\documentclass[12pt]{article}

\newcommand{\bm}{\bar{M}}
\newcommand{\bvarphi}{\bar{\varphi}}
\newcommand{\bv}{\bar{V}}
\newcommand{\hv}{\hat{V}}
\newcommand{\tv}{\tilde{V}}

\title{
MORSE POTENTIAL AND ITS RELATIONSHIP WITH THE COULOMB IN A
POSITION-DEPENDENT MASS BACKGROUND}
\author{B. BAGCHI and P. S. GORAIN\\
{\small \sl Department of Applied Mathematics, University of Calcutta,} \\
{\small \sl 92 Acharya Prafulla Chandra Road, Kolkata 700 009, India}\\
{\small \sl bbagchi123@rediffmail.com}\\ [7pt] 
C. QUESNE\\
{\small \sl Physique Nucl\'eaire Th\'eorique et Physique Math\'ematique,}
\\ {\small \sl Universit\'e Libre de Bruxelles, Campus de la Plaine CP229,} \\ 
{\small \sl Boulevard~du Triomphe, B-1050 Brussels, Belgium} \\
{\small \sl cquesne@ulb.ac.be}}
\date{ }
\begin{document}
\baselineskip=22pt plus 1pt minus 1pt
%%%%%%%%%%%%%%%%%%%%%%%%%%%%%%%%%%%%%%%%%%%%%%%%%%%%%%%%%%
\maketitle

\begin{abstract} 
We provide some explicit examples wherein the Schr\"odinger equation for the Morse
potential  remains exactly solvable in a position-dependent mass background.
Furthermore, we show how in such a context, the map from the full line $(- \infty, 
\infty)$ to the half line $(0, \infty)$ may convert an exactly solvable Morse potential
into an exactly solvable Coulomb one. This generalizes a well-known property of
constant-mass problems.
\end{abstract}

\noindent
Running head: Morse and Coulomb Potentials


%
%========================================================================
%
\newpage
Morse potential plays an important physical role in quantum mechanics (see,
e.g.,~\cite{pauling}). It is an exactly solvable potential and is of much use in
spectroscopic applications. It is also known to have a connection with the Coulomb
potential under a coordinate transformation --- a feature well exploited \cite{haymaker,
lahiri} to describe supersymmetry for the hydrogen atom.\par
%
%------------------------------------------------------------------------------------
% 
In this note we report on certain characteristics of a one-dimensional Morse potential in a
position-dependent mass (PDM) background. There has been considerable interest in
PDM problems for some time \cite{dekar98, dekar99, milanovic, plastino99, plastino00,
dutra00, dutra03, roy02, roy05, koc02, koc03a, koc03b, alhaidari, gonul02a,
gonul02b, gonul05, bagchi04a, bagchi04b, bagchi05a, bagchi05b, yu04a, yu04b, chen,
dong, cq06}. Indeed there exists a wide variety of situations in which PDM is of utmost
relevance \cite{weisbuch}. PDM also holds out to deformations in the quantum canonical
commutation relations or curvature of the underlying space \cite{cq04}. Furthermore, it
has recently been observed that there exists a whole class of Hermitian PDM Hamiltonians
which, to lowest order of perturbation theory, have a correspondence with
pseudo-Hermitian Hamiltonians
\cite{bagchi06}.\par
%
%------------------------------------------------------------------------------
%  
In a series of papers \cite{bagchi04a, bagchi04b, bagchi05a, bagchi05b}, we extensively
discussed the solutions of the one-dimensional Schr\"odinger equation with PDM in the
kinetic energy operator. We explored, in particular, its matching with the
coordinate-transformed constant-mass version and the consequent decoupling of the
ambiguity parameters appearing in the effective potential. We also looked into the viable
choices of the mass function (see also \cite{dutra00, dutra03}).\par
%
%-----------------------------------------------------------------------------
%
Here we show that an exponential choice for the PDM allows one to transform the
corresponding Morse Hamiltonian into the constant-mass problem. The eigenfunctions of
the latter being known, it is a simple exercise to determine those of the former. On
the other hand, a mapping to the half line $(0, \infty)$ results in the expected conversion
of the full line $(- \infty, \infty)$ Morse into the Coulomb. We then demonstrate how to
identify the Coulomb potential in the presence of PDM. We also provide estimates of the
potential parameter, modified angular momentum quantum number and energy eigenvalues
in such a setting.\par
%
%================================================
% 
The one-dimensional effective PDM Hamiltonian is given by \cite{vonroos}
\begin{equation}
  H_{\rm eff} = - \frac{d}{dx} \frac{1}{M(x)} \frac{d}{dx} + V_{\rm eff}(x),
  \label{eq:Heff}
\end{equation}
where $V_{\rm eff}(x)$ contains the given potential $V(x)$:
\begin{equation}
  V_{\rm eff}(x) = V(x) + \frac{1}{2} (\beta + 1) \frac{M''}{M^2} - [\alpha (\alpha +
  \beta + 1) + \beta + 1] \frac{M^{\prime2}}{M^3}.  \label{eq:Veff}
\end{equation}
In (\ref{eq:Veff}), $\alpha$, $\beta$ are the ambiguity parameters and primes stand for
the derivatives with respect to $x$. We use the dimensionless form $M(x)$ of the mass
function $m(x) = m_0 M(x)$ and adopt units such that $\hbar = 2m_0 = 1$.\par
%
%-------------------------------------------------------------------
%
The Schr\"odinger equation corresponding to (\ref{eq:Heff}) is
\begin{equation}
  \left(- \frac{1}{M} \frac{d^2}{dx^2} + \frac{M'}{M^2} \frac{d}{dx} + V_{\rm eff} -
  \epsilon\right) \varphi(x) =0,  \label{eq:SE}
\end{equation}
where the first-derivative term can be removed by the transformation $\varphi =
\sqrt{M}\, \psi$:
\begin{equation}
  \left(- \frac{1}{M} \frac{d^2}{dx^2} + \frac{3}{4} \frac{M^{\prime2}}{M^3}
  - \frac{1}{2} \frac{M''}{M^2} + V_{\rm eff} - \epsilon\right) \psi(x) =0.
  \label{eq:SEbis}
\end{equation}
\par
%
%---------------------------------------------------------------------------
%
Let us set for $M(x)$ and $V(x)$ the following forms
\begin{equation}
  M(x) = e^{-2x}, \qquad V(x) = V_0 e^{2x} - B (2A+1) e^x,  \label{eq:V-Morse}
\end{equation}
where $V_0$, $A$ and $B$ are coupling parameters. It then follows from
(\ref{eq:SEbis}) that 
\begin{equation}
  \left\{- \frac{d^2}{dx^2} + U(x) + V_0 - [4\alpha (\alpha + \beta +1) + 2(\beta + 1)
  - 1]\right\} \psi(x) = 0,
\end{equation}
where $U(x) = B^2 e^{-2x} - B (2A+1) e^{-x}$ and we have put $\epsilon = - B^2$ for
notational reasons. We thus find that the Morse potential $V(x)$ with an exponential PDM
can be transformed into the standard $U(x)$ with constant mass.\footnote{Actually, for
some appropriate choices of parameters, $U(x)$ coincides with the potential proposed by
Morse \cite{morse}, while $V(x)$ is an alternative (often used) form obtained by applying
the parity transformation $x \to -x$.} With $V_0 = (A-n)^2 + [4\alpha (\alpha + \beta
+1) + 2\beta + 1]$, the energy eigenvalues of the latter can be cast as \cite{cooper}
\begin{equation}
  E_n = - (A-n)^2, \qquad n=0, 1, \ldots, n_{\rm max}\ (n_{\rm max} < A).
\end{equation}
Further, the corresponding wavefunctions can be expressed as
\begin{equation}
  \psi_n(x) \propto y^{A-n} e^{- \frac{1}{2} y} L^{(2A-2n)}_n(y), \qquad y = 2B e^{-x}.
\end{equation}
\par
%
%--------------------------------------------------------------------------
%
It is now straightforward to derive the wavefunctions of (\ref{eq:SE}) by employing the
relation $\varphi = e^{-x} \psi$. We obtain
\begin{equation}
  \varphi_n(x) \propto y^{A+1-n} e^{- \frac{1}{2} y} L^{(2A-2n)}_n(y).
\end{equation}
One can see $\varphi_n(x)$ to be square integrable on $(- \infty, \infty)$. Moreover for
$x \to \infty$, the condition $|\varphi_n(x)|^2 / \sqrt{M(x)} \sim e^{-(2A+1-2n)x}
\to 0$ is fulfilled which, as has been established elsewhere \cite{bagchi05a}, is
appropriate for a vanishing mass at $x \to \infty$.\par
%
%---------------------------------------------------------------------------------
%
It must be stressed that other choices of mass functions can be made which work well for
the Morse potential. It is also interesting to observe that the following form for $M(x)$,
\begin{equation}
  M(x) = \frac{1}{(1 + \kappa e^x)^2} \qquad (\kappa > 0),
\end{equation}
renders both $V(x)$ and $V_{\rm eff}(x)$ Morse-like. Indeed plugging in $V(x)$ from
(\ref{eq:V-Morse}) (where we have reset $V_0 = B^2$) gives
\begin{equation}
  V_{\rm eff}(x) = B^{\prime2} e^{2x} - B' (2A'+1) e^x,
\end{equation}
where $B^{\prime2} = B^2 - 2 [2\alpha (\alpha + \beta + 1) + \beta + 1] \kappa^2$
and $B' (2A'+1) = B (2A+1) + (\beta+1) \kappa$. Notice that like $V(x)$, $V_{\rm eff}
(x)$ is also Morse but has the coefficients scaled. In terms of the new parameters $A'$,
$B'$, the energy eigenvalues turn out to be
\begin{equation}
  \epsilon_n = - \frac{1}{4} \left(\frac{2B' (2A'+1) - [(2n+1) \Delta + 2(n^2 + n +1)
  \kappa]}{\Delta + (2n+1) \kappa}\right)^2  \label{eq:epsilon}
\end{equation}
in which $\Delta = 2 \sqrt{B^{\prime2} + \kappa^2}$. The corresponding wavefunctions
$\varphi_n(x)$ can also be determined. For the ground state, for instance, one finds the
function
\begin{eqnarray}
  \varphi_0(x) & \propto & (1 + \kappa e^x)^{\frac{\lambda}{\kappa} - \mu -
        \frac{1}{2}} e^{\mu x}, \nonumber \\ 
  \lambda & = & - \frac{1}{2} (\Delta + \kappa), \qquad \mu = \frac{1}{2}
        \left(\frac{2B' (2A'+1) - \kappa}{\Delta + \kappa} - 1\right),  \label{eq:gs}
\end{eqnarray}
which is square integrable on $(- \infty, \infty)$ and satisfies the condition 
$|\varphi_0(x)|^2 / \sqrt{M(x)} \to 0$ for $x \to \infty$, as it should be.\par
%
%============================================
%
We now turn to the Coulomb problem.\par
%
%---------------------------------------------------------------------
%
A coordinate transformation $x = \ln r$ transforms a full line $(- \infty, \infty)$ to a half
line $(0, \infty)$. Making this change of variable, Eq.~(\ref{eq:SE}) gets modified to the
form
\begin{equation}
  \left[- \frac{1}{\bm} \frac{d^2}{dr^2} + \frac{1}{\bm} \left(\frac{\dot{\bm}}{\bm}
  - \frac{1}{r}\right) \frac{d}{dr} + \frac{1}{r^2} \left(\hv_{\rm eff} - \epsilon_n\right)
  \right] \bvarphi_n(r) = 0,
\end{equation}
where we have denoted
\begin{eqnarray}
  \bm(r) & = & M(x(r)) = \frac{1}{(1 + \kappa r)^2}, \qquad \bvarphi_n (r) = \varphi_n
         (x(r)), \nonumber \\
  \hv_{\rm eff}(r) & = & V_{\rm eff}(x(r)) = B^{\prime2} r^2 - B' (2A'+1) r,
         \label{eq:mbar}   
\end{eqnarray}
and indicated by an overhead dot a derivative with respect to $r$.\par
%
%--------------------------------------------------------------------------
%
A further substitution $\bvarphi(r) = \sqrt{r}\, \chi(r)$ results in the three-dimensional
form of the PDM radial Schr\"odinger equation
\begin{equation}
  \left[- \frac{1}{\bm} \frac{d^2}{dr^2} + \frac{1}{\bm} \left(\frac{\dot{\bm}}{\bm}
  - \frac{2}{r}\right) \frac{d}{dr} + \frac{l(l+1)}{\bm r^2} + \bv_{\rm eff} - E \right]
  \chi_{nl}(r) = 0,  \label{eq:radial-SE}
\end{equation}
in which $l$ is the angular momentum quantum number and $\bv_{\rm eff}(r)$ is given
by
\begin{equation}
  \bv_{\rm eff}(r) = - \frac{1}{\bm} \left(- \frac{1}{2r} \frac{\dot{\bm}}{\bm} +
  \frac{\left(l + \frac{1}{2}\right)^2}{r^2}\right) + \frac{1}{r^2} \left(\hv_{\rm eff}
  - \epsilon_n\right) + E.
\end{equation}
Note that Eq.~(\ref{eq:radial-SE}) could also be arrived at if we generalized
(\ref{eq:Heff}) to three dimensions \cite{cq06} and carried out the usual separation of
variables in the corresponding Schr\"odinger equation.\par
%
%--------------------------------------------------------------------------
%
Actually we can convert (\ref{eq:radial-SE}) to the typical one-dimensional form
(\ref{eq:Heff}), namely
\begin{equation}
  \left(- \frac{d}{dr} \frac{1}{\bm} \frac{d}{dr} + \tv_{\rm eff} - E \right) \xi_{nl}(r) =
  0,
\end{equation}
if we set $\chi(r) = \frac{1}{r} \xi(r)$ and define $\tv_{\rm eff}(r)$ as
\begin{equation}
  \tv_{\rm eff}(r) = \bv_{\rm eff} - \frac{\dot{\bm}}{\bm^2 r} + \frac{l(l+1)}{\bm r^2}.
\end{equation}
\par
%
%----------------------------------------------------------------------------
%
{}For the choice of mass function in (\ref{eq:mbar}), $\tv_{\rm eff}(r)$ becomes
\begin{equation}
  \tv_{\rm eff}(r) = - \frac{B' (2A'+1) + \frac{\kappa}{2}}{r} - \frac{\epsilon_n +
  \frac{1}{4}}{r^2} + E + B^{\prime2} + \frac{3}{4} \kappa^2.
\end{equation}
\par
%
%--------------------------------------------------------------------
% 
We conclude that the function $\xi_{nl}(r)$ satisfies the equation
\begin{equation}
  \left(- \frac{d}{dr} \frac{1}{\bm} \frac{d}{dr} - \frac{2Z}{r} + \frac{l(l+1)}{r^2}
  - \frac{1}{4} \kappa^2 - \lambda_{nl} \right) \xi_{nl}(r) = 0  \label{eq:Coulomb-SE}
\end{equation}
and may be interpreted to be the one satisfied by the Coulomb potential $- \frac{2Z}{r}$
in the presence of PDM. In (\ref{eq:Coulomb-SE}), we defined
\begin{equation}
  2Z = B' (2A'+1) - \frac{\kappa}{2}, \qquad l(l+1) = - \epsilon_n - \frac{1}{4}, \qquad
  \lambda_{nl} = - B^{\prime2} - \kappa^2.  \label{eq:Coulomb-Morse}
\end{equation}
\par
%
%-------------------------------------------------------------------------
%
In the constant-mass case ($\kappa = 0$), we have $A'=A$, $B'=B$, $\epsilon_n = -
(A-n)^2$. On using (\ref{eq:Coulomb-Morse}), we get
\begin{equation}
  Z = B \left(A + \frac{1}{2}\right), \qquad l = A - n - \frac{1}{2}, \qquad
  \lambda_{nl} = - \frac{Z^2}{(n+l+1)^2},  \label{eq:constant}
\end{equation}
and so Eq.~(\ref{eq:Coulomb-SE}) coincides with the standard radial equation of the
Coulomb potential, where $\lambda_{nl}$ denotes the energy eigenvalues.\par
%
%-------------------------------------------------------------------------
%
{}Finally, from (\ref{eq:Coulomb-Morse}) we can obtain solutions for $Z$, $l$ and the
eigenvalues $\lambda_{nl}$ as follows:
\begin{eqnarray}
  Z & = & B' \left(A' + \frac{1}{2}\right) - \frac{\kappa}{4}, \nonumber \\ 
  l & = & \frac{2B' (2A'+1) - [2(n+1) \Delta + (2n^2 + 4n +3) \kappa]}{2[\Delta +
       (2n+1) \kappa]}, \nonumber \\
  \lambda_{nl} & = & - \left(\frac{2Z - [n^2 + (l+1)(2n+1)] \kappa}{2(n+l+1)}\right)^2,
       \label{eq:l-lambda}
\end{eqnarray}
where we have used $\epsilon_n$ given by (\ref{eq:epsilon}). These are to be
interpreted as modified expressions of the parameter $Z$, the angular momentum
quantum number and the eigenvalues in the presence of PDM.\par
%
%--------------------------------------------------------------------------------
%
The corresponding wavefunctions can be easily found from the relation $\xi_{nl}(r) =
\sqrt{r}\, \bvarphi_n(r)$. From (\ref{eq:gs}), for instance, we get
\begin{equation}
  \xi_{0l}(r) \propto r^{\mu + \frac{1}{2}} (1 + \kappa r)^{\frac{\lambda}{\kappa} -
  \mu - \frac{1}{2}},
\end{equation}
which can be rewritten as
\begin{equation}
  \xi_{0l}(r) \propto r^{l+1} (1 + \kappa r)^{-\left(\frac{Z}{(l+1)\kappa} + l +1\right)}
\end{equation}
as a consequence of (\ref{eq:gs}), (\ref{eq:Coulomb-Morse}) and (\ref{eq:l-lambda}). It
can be easily checked that such a transformed wavefunction is square integrable on $(0,
\infty)$ and fulfils the condition $|\xi_{0l}(r)|^2 / \sqrt{\bm(r)} \to 0$ for $r \to
\infty$.\par
%
%-----------------------------------------------------------------------
%
As a final comment, it is worth observing that nonnegative integer values of $l$ in
(\ref{eq:l-lambda})\footnote{The requirement of integer $Z$ values may be relaxed
since in atomic physics an effective parameter $Z_{\rm eff}$ is often used
\cite{pauling}.} are associated with some specific choices for the Morse potential and
mass parameters $A'$, $B'$, $\kappa$. This, however, is by no way a new feature due to
the PDM environment since it can be seen from (\ref{eq:constant}) that a similar
restriction exists for $A$ and $B$ in the constant-mass case.\par
%
%============================================================
%
\section*{Acknowledgments}

B.B. gratefully acknowledges the support of the National Fund for Scientific Research
(FNRS), Belgium, and the warm hospitality at PNTPM, Universit\'e Libre de Bruxelles, where
this work was carried out. P.S.G. thanks the Council of Scientific and Industrial Research
(CSIR), New Delhi, for the award of a fellowship. C.Q. is a Research Director of the National
Fund for Scientific Research (FNRS), Belgium.\par
%
%=========================================================
%


\newpage
\begin{thebibliography}{99}

\bibitem{pauling}L.\ Pauling and E.\ B.\ Wilson, Jr., {\sl Introduction to Quantum
Mechanics with Applications to Chemistry} (Mc Graw-Hill, New York, 1935).

\bibitem{haymaker} R.\ W.\ Haymaker and A.\ R.\ P.\ Rau, {\sl Am.\ J.\ Phys.} {\bf 54},
928 (1986).

\bibitem{lahiri} A.\ Lahiri, P.\ K.\ Roy and B. Bagchi, {\sl J.\ Phys.} {\bf A20}, 3825
(1987).

\bibitem{dekar98} L.\ Dekar, L.\ Chetouani and T.\ F.\ Hammann, {\sl J.\ Math.\ Phys.}
{\bf 39}, 2551 (1998). 

\bibitem{dekar99} L.\ Dekar, L.\ Chetouani and T.\ F.\ Hammann, {\sl Phys.\ Rev.} {\bf
A59}, 107 (1999).

\bibitem{milanovic} V.\ Milanovi\'c and Z.\ Ikoni\'c, {\sl J.\ Phys.} {\bf A32}, 7001
(1999).

\bibitem{plastino99} A.\ R.\ Plastino, A.\ Rigo, M.\ Casas, F.\ Garcias and A.\ Plastino,
{\sl Phys.\ Rev.} {\bf A60}, 4318 (1999).

\bibitem{plastino00}  A.\ R.\ Plastino, A.\ Puente, M.\ Casas,
F.\ Garcias and A.\ Plastino, {\sl Rev.\ Mex.\ Fis.} {\bf 46}, 78 (2000).

\bibitem{dutra00} A.\ de Souza Dutra and C.\ A.\ S.\ Almeida, {\sl Phys.\ Lett.} {\bf
A275}, 25 (2000).

\bibitem{dutra03} A.\ de Souza Dutra, M.\ Hott and C.\ A.\ S.\ Almeida, {\sl
Europhys.\ Lett.} {\bf 62}, 8 (2003).

\bibitem{roy02} B.\ Roy and P.\ Roy, {\sl J.\ Phys.} {\bf A35}, 3961 (2002).

\bibitem{roy05} B.\ Roy and P.\ Roy, {\sl Phys.\ Lett.} {\bf A340}, 70 (2005).

\bibitem{koc02} R.\ Ko\c c, M.\ Koca and E.\ K\"orc\"uk, {\sl J.\ Phys.} {\bf A35}, L527
(2002).

\bibitem{koc03a} R.\ Ko\c c and M.\ Koca, {\sl J.\ Phys.} {\bf A36}, 8105 (2003).

\bibitem{koc03b} R.\ Ko\c c and H.\ T\"ut\"unc\"uler, {\sl Ann.\ Phys.\ (Leipzig)} {\bf
12}, 684 (2003).

\bibitem{alhaidari} A.\ D.\ Alhaidari, {\sl Phys.\ Rev.} {\bf A66}, 042116 (2002). 

\bibitem{gonul02a} B.\ G\"on\"ul, B.\ G\"on\"ul, D.\ Tutcu and O.\ \"Ozer, {\sl Mod.\
Phys.\ Lett.} {\bf A17}, 2057 (2002). 

\bibitem{gonul02b} B.\ G\"on\"ul, O.\ \"Ozer, B.\ G\"on\"ul and
F.\ \"Uzg\"un, {\sl Mod.\ Phys.\ Lett.} {\bf A17}, 2453 (2002).

\bibitem{gonul05} B.\ G\"on\"ul and M.\ Ko\c cak, {\sl Chin.\ Phys.\ Lett.} {\bf 22},
2742 (2005).

\bibitem{bagchi04a} B.\ Bagchi, P.\ Gorain, C.\ Quesne and R.\ Roychoudhury, {\sl
Mod.\ Phys.\ Lett.} {\bf A19}, 2765 (2004).

\bibitem{bagchi04b} B.\ Bagchi, P.\ Gorain, C.\ Quesne and R.\ Roychoudhury, {\sl Czech.\
J.\ Phys.} {\bf 54}, 1019 (2004).

\bibitem{bagchi05a} B.\ Bagchi, A.\ Banerjee, C.\ Quesne and V.\ M.\ Tkachuk, {\sl
J.\ Phys.} {\bf A38}, 2929 (2005).

\bibitem{bagchi05b} B.\ Bagchi, P.\ Gorain, C.\ Quesne and R.\ Roychoudhury, {\sl
Europhys.\ Lett.} {\bf 72}, 155 (2005).

\bibitem{yu04a} J.\ Yu, S.-H.\ Dong and G.-H.\ Sun, {\sl Phys.\ Lett.} {\bf A322}, 290
(2004).

\bibitem{yu04b} J.\ Yu and S.-H.\ Dong, {\sl Phys.\ Lett.} {\bf A325}, 194 (2004).

\bibitem{chen} G.\ Chen and Z.\ Chen, {\sl Phys.\ Lett.} {\bf A331}, 312 (2004).

\bibitem{dong} S.-H.\ Dong and M.\ Lozada-Cassou, {\sl Phys.\ Lett.} {\bf A337}, 313
(2005).

\bibitem{cq06} C.\ Quesne, {\sl Ann.\ Phys.\ (N.\ Y.)} {\bf 321}, 1221 (2006).

\bibitem{weisbuch} C.\ Weisbuch and B.\ Vinter, {\sl Quantum Semiconductor
Heterostructures} (Academic Press, New York, 1997).

\bibitem{cq04} C.\ Quesne and V.\ M.\ Tkachuk, {\sl J.\ Phys.} {\bf A37}, 4267 (2004).

\bibitem{bagchi06} B.\ Bagchi, C.\ Quesne and R.\ Roychoudhury, {\sl J.\ Phys.} {\bf
39}, L127 (2006).

\bibitem{vonroos} O.\ von Roos, {\sl Phys.\ Rev.} {\bf B27}, 7547 (1983).

\bibitem{morse} P.\ M.\ Morse, {\sl Phys.\ Rev.} {\bf 34}, 57 (1929).

\bibitem{cooper} F.\ Cooper, A.\ Khare and U.\ Sukhatme, {\sl Phys.\ Rep.} {\bf 251},
267 (1995).

\end {thebibliography}
\end{document}